\documentclass[fleqn,twoside]{article}
\usepackage{espcrc2}

\usepackage{graphicx}
\usepackage[figuresright]{rotating}

\newcommand{\babar}{BaBar}
\newcommand{\el}{$E_\ell$}
\newcommand{\mx}{$M_X$}
\newcommand{\mxqsq}{$(M_X, q^2)$}
\newcommand{\elsmax}{$(E_\ell, s^{max})$}
\newcommand{\smax}{$s^{max}$}
\newcommand{\pplus}{$P^+$}
\newcommand{\Vub}{$|V_{ub}|$}
\newcommand{\Vcb}{$|V_{cb}|$}
\newcommand{\B}{$B$}

\newcommand{\gev}{$GeV$}

\newcommand{\mb}{$m_b$}
\newcommand{\mc}{$m_c$}
\newcommand{\btoxlnu}{$B\rightarrow X_u \ell \nu_{\ell}$}
\newcommand{\btoxclnu}{$B\rightarrow X_c \ell \nu_{\ell}$}

\newcommand{\AmS}{{\protect\the\textfont2
  A\kern-.1667em\lower.5ex\hbox{M}\kern-.125emS}}

\title{\Vub\ extraction using the Analytic Coupling model}

\author{U.~Aglietti\address{ Dip.\ Fis., Univ.\ di Roma I ``La Sapienza'' \& INFN Roma, 
Roma, Italy },
F.~Di Lodovico\address{Queen Mary, University of London,  Dep. of Phys., London, UK}
G.~Ferrera\address{Dip. Fis., Univ.\ di Firenze \& INFN Firenze, Sesto Fiorentino, Firenze, Italy},
G.~Ricciardi\address{Dip.\ Scienze Fis., Univ.\ di Napoli ``Federico II'' \& INFN Napoli, Napoli, Italy}.}
\begin{document}

\begin{abstract}
We summarize the main characteristics and recent results on \btoxlnu\ decays
of a model based on soft--gluon resummation and an analytic time--like QCD coupling.
\vspace{1pc}
\end{abstract}

\maketitle

\section{Introduction}

By comparing various spectra in the semileptonic inclusive \B\ decays $B
\, \rightarrow \, X_u \, \, \ell \, \, \nu_{\ell},$
where $\ell$ is a fixed lepton species ($\ell=e,\mu$) and $X_u$
the fragmented $u$ quark, with the  predictions
of a model including non--perturbative corrections to soft--gluon
dynamics through an effective QCD coupling~\cite{model}, 
we obtain a value for the \Vub\
Cabibbo--Kobayashi--Maskawa (CKM) matrix element~\cite{ckm,grinstein} 
$|V_{ub}| \, = \, ( \, 3.76 \, \pm \, 0.13 \, \pm \, 0.22 \, ) \times 10^{-3} \, $
being the errors experimental and theoretical, respectively.
The model involves the insertion,
inside standard threshold resummation formulae, of an effective
QCD coupling $\tilde{\alpha}_S(k^2)$, based on an analyticity
requirement and resumming absorptive effects in gluon
cascades~\cite{shirkov}. By construction, $\tilde{\alpha}_S(k^2)$
has no Landau pole and saturates at small 
scales:
\begin{eqnarray} \lim_{k^2
\rightarrow 0} \tilde{\alpha}_S(k^2) \, = \, \frac{1}{\beta_0} \, \approx
\, \mathcal{O}(1) \, . 
\nonumber
\end{eqnarray} 
This model, which has no free parameters, describes
\B--meson fragmentation data at the $Z^0$ peak rather well~\cite{acf}, where
--- unlike $B$ decays --- 
accurate data are available and there is no uncertainty
coming from the CKM matrix elements. 

In the following, we describe the phenomelogical
model used, then the extraction of \Vub, 
and finally the results, followed by the conclusions.

\section{Threshold resummation with an effective coupling}
\label{Threshold--Resummation}

Factorization and resummation of threshold logarithms in
semileptonic decays leads to an expression for the
triple--differential distribution, the most general distribution,
of the following form~\cite{ugo2001}:
\vspace*{-0.05cm}
\begin{eqnarray}
\frac{1}{\Gamma} \frac{d^3\Gamma}{dx dw du} \, = \,
C[x, w; \alpha_S(Q)] \, \sigma[u; Q] \,  \,
\nonumber \\
+D[x,u,w; \alpha_S(Q)] \, ,
\nonumber
\end{eqnarray}
where:
\vspace*{-0.1cm}
\begin{eqnarray}
x \, = \, \frac{2 E_\ell}{m_b}, ~~~
w \, = \, \frac{Q}{m_b}, ~~~
u \, = \, \frac{1 - \sqrt{1 - (2m_X/Q)^2} }{1 + \sqrt{1 - (2m_X/Q)^2} }
\nonumber
\end{eqnarray}
and the hard scale $Q = 2 E_X$, with $E_\ell$, $E_X$ and $m_X$ being the charged 
lepton energy, the total hadron energy and the hadron mass, respectively.
$\Gamma=\Gamma(\alpha_S)$ is the inclusive width of decay $B
\, \rightarrow \, X_u \, \, \ell \, \, \nu_{\ell}$. 
Furthermore, $C[x,w; \alpha_S]$ is a short--distance, process dependent hard factor;
$\sigma[u; Q]$ is the universal QCD form factor for heavy--to--light transitions,
resumming to any order in $\alpha_S$ the series of logarithmically enhanced terms
to some logarithmic accuracy;
$D[x,u,w;\alpha_S]$ is a short--distance, process dependent, remainder function,
vanishing in the threshold region $u \rightarrow 0$ and in lowest--order in $\alpha_S$.
The heavy flavor decay form factor has an exponential form in Mellin moments
$N$--space~\cite{cattren}. We apply a change of renormalization scheme for the coupling
constant $\alpha_S \rightarrow \tilde{\alpha}_S$. 
The QCD form factor $\sigma[u; Q]$ has been numerically computed 
for different values of $\alpha_S(m_Z)$ in~\cite{model}.

Threshold suppression --- the main theoretical ingredient
for the measurement of $|V_{ub}|$ ---  is represented by the factor
\begin{eqnarray}
\label{defW}
W(a,b) \equiv \frac{
\Gamma\left[ B \rightarrow X_u \ell \nu_{\ell} , p \in (a, b)
\right]} {\Gamma\left[B \, \rightarrow \, X_u \, \ell \, \nu_{\ell}\right]}
\nonumber \\
= \int_{a<p(x,w,u)<b} \frac{1}{\Gamma} \frac{d^3\Gamma}{dx dw du}dx dw du 
\, \le \, 1 \, .
\end{eqnarray}
where $W(a,b) = 1$ when integrated over all the phase--space of $x, \, w$ and $u$.
For threshold resummed spectra of $B \rightarrow X_u \, \ell\, \nu_{\ell}$ decays at
next--to--leading order see Ref.~\cite{noi}.

\section{$ |V_{ub}| $ extraction}
\label{meth}

Experimentally, given a kinematical variable $p$,
such as for example the energy \el\ of the charged lepton, one
measures the number of \B's decaying semileptonically to $X_u$ 
with $p$ in some interval $(a,b)$, divided by the total number of produced
\B's (decaying into any possible final state): 
\begin{equation} \label{br} {\cal B}\left[ p \in (a,b) \right] \, \equiv
\, \frac{ N [ B \, \rightarrow \, X_u \, \ell \, \nu_{\ell} , ~ p \in (a,b) ] } {
N\left[ B \rightarrow {\rm (anything)} \right] } \, . \end{equation} 
This branching ratio can be written as:
\begin{equation} 
\label{baseq} 
{\cal B}\left[ p \in (a, b) \right] \,
= \, \frac{ {\cal B}_{SL} }{1 \, + \, {\cal R}_{c/u} } \, W(a,b) \, ,
\end{equation} 
where we have defined the semileptonic branching ratio:
\begin{eqnarray}
{\cal B}_{\rm SL} \, \equiv \, \frac{\Gamma(B \, \rightarrow \, X_c \, \ell
\, \nu_{\ell}) \, + \, \Gamma(B \, \rightarrow \, X_u \, \ell \,
\nu_{\ell})}{\Gamma\left[B \, \rightarrow \, {\rm (anything)} \right] } \nonumber\end{eqnarray}
and the ratio of $(b \rightarrow c)/(b \rightarrow u)$ semileptonic widths: \begin{equation}
\label{defH} {\cal R}_{c/u} \, \equiv \, \frac{ \Gamma( B \, \rightarrow
\, X_c \, \ell \, \nu_{\ell} ) }{ \Gamma( B \, \rightarrow \, X_u \, \ell \, \nu_{\ell} )
} \, . \end{equation} Since ${\cal B}_{\rm SL}$ is rather well measured, we use
the experimental determination~\cite{PDG}: \begin{eqnarray} {\cal B}_{\rm SL} \, = \, 0.1066 \, \pm \, 0.0020
\, . \nonumber
\end{eqnarray} 

With this method there is no $m_b^5$ dependence (with the related
uncertainty) which would appear using the theoretical expression of 
$\Gamma( B \, \rightarrow \, X_u \, \ell \, \nu_{\ell})$,
because one has to compute only
the ratio of widths ${\cal R}_{c/u} $ and not the absolute widths.
The semileptonic $b \rightarrow c$ width is written as: 
\begin{eqnarray}
\Gamma( B \, \rightarrow \, X_c \, \ell \, \nu_{\ell} ) \, = 
\frac{G_F^2 \,
m_b^5 \, |V_{cb}|^2 }{192 \pi^3} \, \nonumber \\ \times I(\rho) \, F(\alpha_S) \,
G(\alpha_S,\rho) \, , \end{eqnarray} where $\rho \, \equiv \,
\frac{m_c^2}{m_b^2} \, \approx 0.1 $. 
The function
$I(\rho)$ accounts for the suppression of phase--space because of
$m_c \ne 0$~\cite{nir} : 
\begin{eqnarray} 
I(\rho) \, = \, 1 - 8 \rho + 12
\rho^2 \log \frac{1}{\rho} + 8 \rho^3 - \rho^4 \, . 
\nonumber
\end{eqnarray} 
Note that there is an (accidental) strong dependence on the charm mass
$m_c$, because of the appearance of a large factor in the leading
term in $\rho$, namely $- \, 8$. As far as inclusive quantities
are concerned, the largest source of theoretical error comes
indeed from the the uncertainty in $\rho$. Most of the dependence
is actually on the difference $m_b - m_c$, which can be estimated
quite reasonably with the Heavy Quark Effective Theory (HQET).
Finally, the factor $G(\alpha_S,\rho)$ contains
corrections suppressed by powers of $\alpha_S$ and
$\rho$: \begin{eqnarray} G(\alpha_S,\rho) \, = \, 1 \, + \,
\sum_{n=1}^{\infty} G_n(\rho) \, \alpha_S^n \, , \nonumber \end{eqnarray} with $G_n(0)
= 0$. Note that $G(0,\rho) \, = \, G(\alpha_S,0) \, = \, 1 $. 
By inserting the above expressions for the semileptonic
rates, one obtains for the perturbative expansion of ${\cal
R}_{c/u}$:
\begin{eqnarray} 
{\cal R}_{c/u} 
\, = \, {\cal R}_{c/u} \left( \rho, \alpha_S, |V_{ub}|/|V_{cb}| \right) \, = \,
\nonumber \\
\frac{|V_{cb}|^2}{|V_{ub}|^2} \, I(\rho) \, G(\alpha_S, \rho) \, .
\nonumber
\end{eqnarray} 
This method actually provides a measurement of
the ratio $|V_{ub}|/|V_{cb}|$, but since the error on $|V_{cb}|$ is rather 
small and theoretically well understood, one is basically measuring 
\Vub.\footnote{ The average of 
determinations of \Vcb\ coming from a global fit to the \btoxclnu\ and 
$b\rightarrow s \gamma$ moments in the kinetic and 1S schemes,
in good agreement with each other,
is \Vcb $= ( 41.6 \pm 0.6) \times 10^{-3} $~\cite{hfag,PDG}. }

\subsection{Quark masses}
\label{massebec}

The approach we are following to compute \Vub\ can be subdivided into
two parts: in the first part we compute the triple--differential
distribution and in the second part the \Vub\ value.
As far as the triple differential distribution is concerned,
as the whole process is described in a perturbative framework,
we do not distinguish between the mass of the $B$ meson and the pole
mass of the $b$ quark, i.e. we consistently assume $m_b=m_B$.

Once we compute the ratio \Vub/\Vcb, we use the standard HQET formulas, 
based on the $b$--quark mass. Thus,
the $b$--quark mass is introduced in our formulation of the ratio.
Since quarks are confined inside observable hadrons, 
their masses cannot be directly measured and their values 
are biased by the selected theoretical framework.
We have performed the calculation  in the $\overline{\mbox{MS}}$
mass scheme. The $\overline{\mbox{MS}}$ masses for the $b$ and the
$c$ quark are taken $\overline{m}_b(\overline{m}_b)= 4.243 \, \pm 0.042 \,
\rm{GeV}$ and $ \overline{m}_c(\overline{m}_c)= 1.25 \, \pm 0.09
\, \rm{GeV}$~\cite{hfag,PDG}, respectively. However, 
we have considered the pole--mass scheme as well 
in order to take
into account the uncertainties coming from a different scheme
definition.

\section{Results}
\label{res}

We calculate \Vub\ for all the experimental analyses. They are
categorized according to the kinematical distribution looked at, where
selection criteria are applied to define the limited phase--space on
which the branching ratio is computed. They are:
the lepton energy (\el), the invariant mass of
the hadron final state (\mx), the light--cone distribution 
(\pplus $\equiv E_X - |\vec{p}_X|$, $E_X$ and $\vec{p}_X$ being the energy
and the magnitude of the 3--momentum of the hadronic system),  
a two dimensional distribution in the
electron energy and \smax,  the maximal \mx$^2$ at fixed $q^2$ and
\el. 
Moreover, as described in Ref.~\cite{model}, we will look only at the range 
where data are not affected by potential 
$b\rightarrow c\, \ell\, \nu_\ell$ background: 2.3~\gev $<$ \el $<$ 2.6~\gev , 
although the method works for lower lepton energy as well.

We compute \Vub\ for each of the analyses starting from the
corresponding partial branching fractions. Then, we determine
the average \Vub\ value using the HFAG methodology~\cite{hfag}.

Table~\ref{tab:global_average} reports the extracted values of
\Vub\ for all the uncorrelated analyses and their corresponding
average. The errors are experimental (i.e. statistical and
systematic) and theoretical, respectively.
The average is:
\begin{eqnarray}
V_{ub} \, = \, ( \, 3.76 \, \pm \, 0.13 \, \pm \, 0.22 \, ) \times 10^{-3} \, ,
\nonumber
\end{eqnarray}
consistent with the measured value of \Vub\ from
exclusive decays~\cite{hfag} and the indirect measurement~\cite{utfit}.

\begin{figure}[!htb]
 \begin{centering}
  \includegraphics[width=0.45\textwidth,totalheight=8cm]{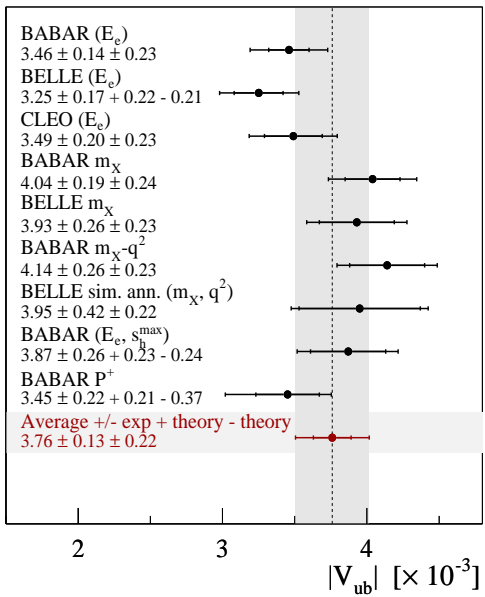}
  \vspace{-0.3in}
  \caption{ \Vub\ values for the uncorrelated analyses and their average.}
  \label{fig:vub}
 \end{centering}
 \end{figure}

The table shows also the
criteria used for the determination of the partial branching ratio
($\Delta{\cal B}$). The theoretical errors are considered
completely correlated among all the experimental analyses, when
performing the average. The
\Vub\ values and the corresponding average are plotted in
Figure~\ref{fig:vub}.

\begin{table*}[!htb]
\begin{center}
\caption{ The first column in the table shows the uncorrelated
analyses, the second column shows the corresponding values of
\Vub, and finally the last column shows the criteria for which
$\Delta{\cal B}$ is available. The final row shows the average
value of \Vub. The errors on the \Vub\ values are experimental and
theoretical, respectively. The experimental error includes both
the statistical and systematic errors.}
\vspace{0.1in}
\begin{tabular}{l|c|c}
\hline\hline
Analysis&   \Vub ($10^{-3}$)           &  $\Delta{\cal B}$ criteria \\
\hline
\babar\ (\el)  \cite{babar-el}  &   3.46$\pm$ 0.14 $^{+0.23}_{-0.23}$  &  \el $ >2.3$\gev\\
Belle (\el)  \cite{belle-el}  &   3.20$\pm$ 0.17 $^{+0.22}_{-0.21}$  &  \el $ >2.3$\gev\\
CLEO  (\el)  \cite{cleo-el}   &   3.49$\pm$ 0.20 $^{+0.23}_{-0.23}$  &  \el $ >2.3$\gev\\
\babar\ (\mx)   \cite{babar-breco}   &   4.04$\pm$ 0.19 $^{+0.24}_{-0.24}$  & \mx $<1.55$~\gev\\
Belle (\mx)   \cite{belle-breco}   &   3.93$\pm$ 0.26 $^{+0.23}_{-0.23}$  & \mx $<1.7$~\gev\\
\babar\ (\mxqsq)   \cite{babar-breco}   &   4.14$\pm$ 0.26 $^{+0.23}_{-0.23}$  & \mx $<1.7$~\gev, $q^2>8$~$\gev^2$ \\
Belle \mxqsq   \cite{belle-ann}   &   3.95$\pm$ 0.42 $^{+0.22}_{-0.22}$  &\mx $<1.7$~\gev, $q^2>8$~$\gev^2$ \\
\babar\ \elsmax \cite{babar-elsmax}&   3.87$\pm$ 0.26 $^{+0.23}_{-0.24}$  & \el $ >2.0$~\gev , \smax$<3.5$~$\gev^2$\\
\babar\ (\pplus)   \cite{babar-breco}   &   3.45$\pm$ 0.22 $^{+0.21}_{-0.37}$  & \pplus $<0.66$~\gev\\
\hline
Average                            &   3.76$\pm$ 0.13 $^{+0.22}_{-0.22}$  &\\
\hline\hline
\end{tabular}
\label{tab:global_average}
\end{center}
\end{table*}

Several sources of theoretical errors have been considered and
shown in Table~\ref{tab:theo_errors}: 
another method, based on the absolute values of the decay rates, to extract 
the value of \Vub, the pole instead of the $\overline{MS}$
mass scheme, the order at which the rate is computed from the exact NLO to the
approximate NNLO~\cite{rit-cza}, the variation of the parameters 
in the computation of \Vub\ within their errors, as given by the
PDG~\cite{PDG}.
What we {\it cannot change} is the modelling of the threshold region,
which is fixed in our model, because it has no free parameters.
That is the factor $W(a,b)$.
The error on the modelling of the threshold region can only be
estimated in an indirect way, by considering different decay spectra, 
in which presumably threshold effects enter in different ways.
Even though our
model on soft--gluon dynamics is formally without free parameters,
we may say that we have constructed it, out of many possibilities,
as a kind of ``expert system''. Once ``trained'' by giving beauty
fragmentation data in input, it should predict reasonable beauty
decay spectra.

\begin{table*}[!htb]
\vspace{-.1in}
\begin{center}
\caption{The first column of the table shows the different
contributions to the theoretical errors, the second column shows
the corresponding variation, and finally the third column shows
the percentage contribution with respect to the \Vub\ value.}
\vspace{0.1in}
\begin{tabular}{l|c|c}
\hline\hline \multicolumn{3}{c}{Theoretical Errors}
\\\hline
Contribution & Variation & Error (\%)
\\\hline
$\alpha_S$ & $0.1176 \pm 0.024$ & $\pm 0.6 \rightarrow 3.5$
\\
\Vcb &$(41.6 \pm 0.6)\times 10^{-3}$ & $\pm 1.4$
\\
\mb\ (\gev) &$4.20 \pm 0.07 $ & $\pm 0.6$
\\
\mc\ (\gev) &$1.25 \pm 0.09 $ & $\pm 4.4$
\\
${\cal B}$ (\btoxlnu)  &$0.1066 \pm 0.0020$& $\pm 1.0$
\\
\Vub\ method && $+0.8$
\\
pole mass (\gev)&
$\begin{array}{c}
4.7<m_b<5.0,\,\,1.47<m_c<1.83\\
3.34<m_b-m_c<3.41
\end{array}$
&$-1.3 \rightarrow  -5.2$
\\
approx. NNLO rate && $+2.0$
\\
\hline\hline
\end{tabular}
\label{tab:theo_errors}
\end{center}
\end{table*}

Table~\ref{tab:partial_averages} shows the \Vub\ averages for
different analysis categories. Note that the \mxqsq\ analyses tend
to have the largest values of \Vub, while the endpoint analyses
the smallest.
The larger value of \Vub\ coming from the analysis of the double
distribution in $(m_X, q^2)$ is expected on qualitative basis.
The lower cut on $q^2$ therefore significantly
reduces the hard scale $Q$ from the ``natural'' value $Q = m_B$, 
where our model has not been tested as it has been constructed to
describe \B--decay spectra having the (maximal) hard scale $Q = m_B$.

\begin{table*}[!htb]
\begin{center}
\vspace{-.1in}
\caption{The table contains the \Vub\ values for several analyses
and the corresponding averages. The errors on the \Vub\ values are
experimental and theoretical, respectively. The experimental error
includes both the statistical and systematic errors.}
\vspace{0.1in}
\begin{tabular}{l|c|c}
\hline\hline \multicolumn{3}{c}{\Vub\ for endpoint analyses ($10^{-3}$)}\\\hline
\babar\ (\el) \cite{babar-el}    &   3.46$\pm$ 0.14 $^{+0.23}_{-0.23}$& \el $>2.3$ \gev \\
Belle (\el) \cite{belle-el}      &   3.20$\pm$ 0.17 $^{+0.21}_{-0.21}$ & \el $>2.3$ \gev\\
CLEO (\el) \cite{cleo-el}        &   3.49$\pm$ 0.20 $^{+0.23}_{-0.23}$ &  \el $>2.3$ \gev\\
\hline Average                   &   3.42$\pm$ 0.15 $^{+0.23}_{-0.22}$  &\\
\hline\hline \multicolumn{3}{c}{\Vub\ for \mx\ analyses ($10^{-3}$)}\\\hline
\babar\ (\mx)   \cite{babar-breco}    &   4.04$\pm$ 0.19 $^{+0.24}_{-0.24}$  &\mx $<1.55$~\gev \\
Belle (\mx)   \cite{belle-breco}      &   3.93$\pm$ 0.26 $^{+0.23}_{-0.22}$ &\mx $<1.7$~\gev \\
\hline Average                        &   4.00$\pm$ 0.16 $^{+0.24}_{-0.23}$ &\\
\hline\hline \multicolumn{3}{c}{\Vub\ for \mxqsq\
analyses ($10^{-3}$)}\\\hline
\babar\ \mxqsq   \cite{babar-breco}  &   4.14$\pm$ 0.26 $^{+0.23}_{-0.23}$ &\mx $<1.7$~\gev, $q^2>8~\gev^2$\\
Belle \mxqsq   \cite{belle-breco}    &   4.21$\pm$ 0.37 $^{+0.23}_{-0.23}$ &\mx $<1.7$~\gev, $q^2>8~\gev^2$\\
Belle \mxqsq   \cite{belle-ann}      &   3.95$\pm$ 0.42 $^{+0.22}_{-0.22}$ &\mx $<1.7$~\gev, $q^2>8~\gev^2$\\
\hline Average                    &   4.13$\pm$ 0.21 $^{+0.23}_{-0.23}$ &\\
\hline\hline \multicolumn{3}{c}{\Vub\ for \pplus\
analyses ($10^{-3}$)}\\\hline
\babar\ (\pplus) \cite{babar-breco}  &   3.45$\pm$ 0.22 $^{+0.21}_{-0.37}$ &$P^+ <0.66$ \\
Belle (\pplus) \cite{belle-breco}    &   3.73$\pm$ 0.32 $^{+0.23}_{-0.29}$&$P^+ <0.66$ \\
\hline Average                       &   3.55$\pm$ 0.19 $^{+0.21}_{-0.23}$ &\\
\hline\hline
\end{tabular}
\label{tab:partial_averages}
\end{center}
\end{table*}

Finally, using the value of $\Delta{\cal B}$ corresponding to the lowest lepton energy cut
for the endpoint analyses, the value of \Vub\ is 3\%\ higher than what we quote
adopting a cut at 2.3~\gev, qualitatively 
consistent with the expectactions due to the observation of a larger number of events
in that region than predicted~\cite{model}.

\section{Conclusions}
\label{concl}

We have analyzed semileptonic $B$ decay data in the framework of a
model for QCD non--perturbative effects based on an effective
time-like QCD coupling, free from Landau singularities. 

Our inclusive measurement of the \Vub\ CKM matrix element is: \begin{eqnarray} |V_{ub}|
\, = \, ( \, 3.76 \, \pm \, 0.13 \, \pm \, 0.22 \, ) \times
10^{-3} . \nonumber  \end{eqnarray}
The errors on the \Vub\ value are experimental
and theoretical, respectively. The experimental error includes
both the statistical and systematic errors.
This value is fully consistent with the determination from exclusive
decays ($|V_{ub}| = (3.51\pm 0.21 ^{+0.66}_{-0.42}) \times  10^{-3}$)~\footnote{The value is obtained from an
average of the Lattice QCD determinations~\cite{hfag}.}
and from an indirect estimate ($|V_{ub}| = ( 3.44 \pm 0.16~\cite{utfit} ) \times 10^{-3} $), whilst
other methods show a discrepancy up to $\approx +2\,\sigma$~\cite{hfag} from those. 
We argue that the main difference of our model with respect to previous ones is a smaller
suppression of the threshold region.


\begin{thebibliography}{9}


\bibitem{model}
 U.~Aglietti, G.~Ferrera and G.~Ricciardi,
  Nucl.\ Phys.\  B {\bf 768} (2007) 85
  [arXiv:hep-ph/0608047].

\bibitem{ckm}
  N.\ Cabibbo, Phys.\ Rev.\ Lett.\ {\bf 10}, 531 (1963);
  M.\ Kobaya\-shi and T.~Maskawa, Prog.\ Theor.\ Phys.\ {\bf 49}, 652 (1973).

\bibitem{grinstein}
For a recent reviw see for example:  B.~Grinstein,
{\it In the Proceedings of 5th Flavor Physics and CP Violation Conference (FPCP 2007), Bled, Slovenia, 12-16 May 2007, pp 005}
  [arXiv:0706.4173 [hep-ph]].


\bibitem{shirkov}
  D.\ V.\ Shirkov and I.\ L.\ Solovtsov,
  Phys.\ Rev.\ Lett.\  {\bf 79} (1997) 1209
  [arXiv:hep-ph/9704333].


\bibitem{acf}
  U.~Aglietti, G.~Corcella and G.~Ferrera,
  Nucl.\ Phys.\  B {\bf 775} (2007) 162
  [arXiv:hep-ph/0610035].


\bibitem{ugo2001}
  U.\ Aglietti,
  Nucl.\ Phys.\  B {\bf 610} (2001) 293
  [arXiv:hep-ph/0104020].



\bibitem{cattren}
  S.~Catani and L.~Trentadue,
  Nucl.\ Phys.\ B {\bf 327} (1989) 323;
  G.~Sterman,
  Nucl.\ Phys.\ B {\bf 281} (1987) 310.



\bibitem{noi}
  U.\ Aglietti, G.\ Ricciardi and G.\ Ferrera,
   Phys.\ Rev.\  D {\bf 74} (2006) 034004
  [arXiv:hep-ph/0507285];
  Phys.\ Rev.\  D {\bf 74} (2006) 034005
  [arXiv:hep-ph/0509095];
  Phys.\ Rev.\  D {\bf 74} (2006) 034006
  [arXiv:hep-ph/0509271].

\bibitem{PDG}
  C.~Amsler {\it et al.}  [Particle Data Group],
  Phys.\ Lett. B {\bf 667} (2008) 1.


\bibitem{nir}
  Y.~Nir,
  Phys.\ Lett.\  B {\bf 221}, 184 (1989).


\bibitem{hfag}
  E.~Barberio {\it et al.},
  arXiv:0808.1297 [hep-ex].
  and online HFAG updates for ICHEP08.


\bibitem{utfit}
  M.~Bona {\it et al.}  [UTfit Collaboration],
  JHEP {\bf 0610} (2006) 081
  [arXiv:hep-ph/0606167].

\bibitem{babar-el} B.~Aubert {\it et al.}  [\babar\ Collaboration], Phys.\ Rev.\ D {\bf 73}, 012006 (2006) [arXiv:hep-ex/0509040].

\bibitem{belle-el}  A.~Limosani {\it et al.}  [Belle Collaboration], Phys.\ Lett.\  B {\bf 621} (2005) 28 [arXiv:hep-ex/0504046].

\bibitem{cleo-el}  A.~Bornheim {\it et al.} [CLEO Collaboration], Phys.\ Rev.\ Lett.\ {\bf 100} (2008) 171802 [arXiv:hep-ex/0202019].

\bibitem{babar-breco} B.~Aubert {\it et al.}  [\babar\ Collaboration], Phys.\ Rev.\ Lett.\ {\bf 88} (2002) 231803 [arXiv:0708.3702 [hep-ex]].

\bibitem{belle-breco} I.~Bizjak {\it et al.}  [Belle Collaboration], Phys.\ Rev.\ Lett.\  {\bf 95} (2005) 241801 [arXiv:hep-ex/0505088].

\bibitem{belle-ann} H.~Kakuno {\it et al.} [Belle Collaboration], Phys.\ Rev.\ Lett.\ {\bf 92} (2004) 101801  [arXiv:hep-ex/0311048].

\bibitem{babar-elsmax} B.~Aubert {\it et al.}  [\babar\ Collaboration], Phys.\ Rev.\ Lett.\ {\bf 95} (2005) 111801, Erratum-ibid.\ {\bf 97} (2006) 019903 [arXiv:hep-ex/0506036].

\bibitem{rit-cza}
  T.~van Ritbergen,
  Phys.\ Lett.\  B {\bf 454} (1999) 353
  [arXiv:hep-ph/9903226];
  A.~Czarnecki and K.~Melnikov,
  Phys.\ Rev.\  D {\bf 59} (1999) 014036
  [arXiv:hep-ph/9804215].


\bibitem{gardi}
E.~Gardi,
  JHEP {\bf 0404} (2004) 049
  [arXiv:hep-ph/0403249];
  J.~R.~Andersen and E.~Gardi,
  JHEP {\bf 0601} (2006) 097
  [arXiv:hep-ph/0509360].


\bibitem{sf}
 I.~I.~Y.~Bigi, M.~A.~Shifman, N.~G.~Uraltsev and A.~I.~Vainshtein,
  Int.\ J.\ Mod.\ Phys.\  A {\bf 9} (1994) 2467
  [arXiv:hep-ph/9312359].


\end{thebibliography}
\end{document}